\documentstyle[preprint,aps,12pt]{revtex}
\tightenlines
\begin{document}
\draft
\hfill\vbox{\baselineskip14pt
            \hbox{\bf PCP-3/ISS 2002}
            \hbox{July 2002}}
\baselineskip20pt
\vskip 0.2cm 
\begin{center}
{\Large\bf Quantum Group Based Theory for Antiferromagnetism 
and Superconductivity: Proof and Further Evidence}
\end{center} 
\begin{center}
\large Sher~Alam$^{1}$, S.~M. Mamun$^{3}$,~T.~Yanagisawa$^{2}$,
M.~O.~Rahman$^{1}$ and J.A.S.~Termizi$^{}$
\end{center}
\begin{center}
$^{1}${\it Photonics, Nat.~Inst.~of~AIST, Tsukuba, Ibaraki 305-8568, Japan}\\
$^{2}${\it Nanoelectronics, Nat.~Inst.~of~AIST, Tsukuba, Ibaraki 
305-8568, Japan}\\
$^{3}${\it  GUAS \& Photon Factory, KEK, Tsukuba, Ibaraki 305, Japan}
$^{4}${\it Department of Physics, Islamia College, Peshawar, NWFP, Pakistan.}
\end{center}
\begin{center} 
\large Abstract
\end{center}
\begin{center}
\begin{minipage}{16cm}
\baselineskip=18pt
\noindent
	Previously one of us presented a conjecture [APF-4 Proceedings]
to model antiferromagnetism and high temperature superconductivity
and their 'unification' by quantum group symmetry rather than 
the corresponding classical symmetry in view of the critique
by Baskaran and Anderson of Zhang's classical SO(5) model.
This conjecture was further sharpened, experimental evidence
and the important role of 1-d systems [stripes] was emphasized
and moreover the relationship between quantum groups and strings 
via WZWN models were given in [Phys. Lett A272, (2000)]. 
In this brief note we give and discuss mathematical proof
of this conjecture, which completes an important part of this
idea, since previously an explicit simple mathematical proof  
was lacking. Moreover an independent calculation [IC/99/2]
which constructs the generators forming SO(5) algebra
not only supports our previous conjecture but provides
a check on our calculations. It is important to note that
in terms of physics that the arbitariness [freedom] of the 
d-wave factor g$^{2}$(k) is tied to quantum group symmetry
whereas in order to recover classical SO(5) one must set
it to unity in an adhoc manner. We intuitively expect
that this freedom may be related to psuedogap behaviour
in cuprates.
   
\end{minipage}
\end{center}
\vfill
\baselineskip=20pt
\normalsize
\newpage
\setcounter{page}{2}
	
\section{Introduction}  
	The parent compounds of HTSC cuprates superconductors
are Antiferromagnetic [AF] Mott insulators. These materials
turn superconducting [SC] upon doping. The existence of 
psuedo-gap [PG] in these materials is yet another interesting
feature. Spin glass behaviour has been reported in some of
the cuprates [such as LSCO]. It is also of interest to test the
notion of quantum phase transitions and the effects of quantum
critical points at higher temperatures in these and related
materials. Thus it is natural to understand
the different states or phases of these materials from a
consistent theoretical construction.

	With some of these points in mind, previously one 
of us presented a conjecture \cite{alam98} to model 
antiferromagnetism and high temperature superconductivity 
and their 'unification' by quantum group symmetry rather than 
the corresponding classical symmetry in view of the critique
by Baskaran and Anderson of Zhang's classical SO(5) model.
This conjecture was further sharpened, experimental evidence
and the important role of 1-d systems [stripes] was emphasized
and moreover the relationship between quantum groups and strings 
via WZWN models were given in  \cite{alam99-1}. 

	There are several ways to implement the quantum group
symmetry, such as starting from Hamiltonian which has
this symmetry. For example it is known that the Hubbard
model at half-filling has exact quantum group symmetry
in 1-d. A generalization of the Hubbard model with
full quantum group symmetry away from half-filling
and 1-d has been given by Schupp \cite{sch97}. 
Another method is to consider the 1-d Hamiltonians
such as Hubbard with quantum group symmetry and use
these to study the 2-d and 3-d HTSC problem via the
string formalism \cite{alam99-1}. In addition one can
formulate schemes of quantum group symmetry breaking.
Yet another approach is to start with a classical groups 
that are relevant to AF and SC embed them in a larger classical
group and then consider the quantum deformation of
this group. Several choices exist \cite{alam99-1}.
The main purpose of this note is to 
consider the choice SO(5)\cite{zhang97}. 

\section{Quantum SO(5) Group}

	It is clear that in order to construct the quantum
SO(5) algebra we must consider the deformation of ordinary
commutation relations. For the bosonic harmonic oscillator
SU(2)$_{q}$ was considered by Biedenharn \cite{bie89}
and Macfarlane \cite{mac89}.  
In the same manner we can deform the anticommutation
relationships of the fermion operators $c$ and $c^{\dagger}$
\begin{eqnarray}
c_{{\bf k},i}^{}c_{{\bf l},j}^{\dagger}
+\left( g({\bf k})g({\bf l})\right) ^{-1}c_{{\bf l},j}^{\dagger}
c_{{\bf k},i}^{} &=&
\left( g^{-2}({\bf k})\right) ^{Q}\delta _{{\bf kl}}\delta _{ij}, 
\nonumber\\   
g({\bf l})c_{{\bf k},i}^{}c_{{\bf l},j}^{}
+g({\bf k})c_{{\bf l},j}^{}c_{{\bf k},i}^{} &=& 0,
\label{q1}
\end{eqnarray}
where $g$ is the deformation function, and $Q$ is defined
such that the following relations hold
\begin{eqnarray}
\left[Q,c_{{\bf k},i}\right] &=& -c_{{\bf k},i}\nonumber\\
\left[Q,c_{{\bf k},i}^{\dagger}\right] 
&=& c_{{\bf k},i}^{\dagger}\nonumber\\
 Q &=& N-M,
\label{q2}
\end{eqnarray}
where N and M are respectively the number of electrons
and number of sites.

To determine $Q$ we set $i=j$ and $\bf{k}=\bf{l}$ 
which reduces Eq.~\ref{q1} to 
\begin{eqnarray}
c_{{\bf k},i}^{}c_{{\bf k},i}^{\dagger}
+\left(g({\bf k})\right) ^{-2}c_{{\bf k},i}^{\dagger}
c_{{\bf k},i}^{} &=&
\left( g^{-2}({\bf k})\right) ^{Q}, 
\nonumber\\   
c_{{\bf k},i}^{}c_{{\bf k},i}^{}
+c_{{\bf k},i}^{}c_{{\bf k},i}^{} &=& 0,
\label{q3}
\end{eqnarray}
and on using Eqs.~\ref{q2} and ~\ref{q3} after a 
short calculation we obtain 
\begin{equation}
Q=\frac{1}{2}{\sum_{\bf k}}\left( g^2({\bf k})\right) ^Q\left(
g^{-2}({\bf p})c_{{\bf k},i}^{\dagger}c_{{\bf k},i}^{}
-c_{{\bf k},i}^{}c_{{\bf k},i}^{\dagger}\right) .
\label{q4}
\end{equation}

	Having obtained the U(1)\footnote{We note that U(1)
is equivalent to rotation group in 2d i.e. SO(2) and SO(3) has
the same Lie Algebra as SU(2)} charge generator $Q$ let us 
incorporate it into a bigger group [here taken to be SO(5)]
 by taking it in conjunction
with  SO(3) spin rotation representing AF \cite{alam99-1}.
The deformed version of rotation generators of SO(3) can be
written in our case as
\begin{eqnarray}
S_1 &=& \frac{1}{2}
\sum_{k} (g({\bf k})^{2})^{Q-1}c_{k,i}^{\dagger}
(\sigma^{x})_{ij}c_{k,j},\nonumber\\
S_2 &=& \frac{1}{2} 
\sum_{k} (g({\bf k})^{2})^{Q-1}c_{k,i}^{\dagger}
(\sigma^{y})_{ij}c_{k,j},\nonumber\\
S_3 &=& \frac{1}{2} 
\sum_{k} (g({\bf k})^{2})^{Q-1}c_{k,i}^{\dagger}
(\sigma^{z})_{ij}c_{k,j}.
\label{4a}
\end{eqnarray}
 
From standard group theory it is known that the 
generators $L_{ab}$ $a,b=1,..,N$ of the SO(N) ``rotation''
groups are $N(N-1)/2$ in number, and satisfy the
following fundamental relation,
\begin{equation}
\left[L_{ab},L_{cd}\right] = i(\delta_{bd}L_{ac}
+\delta_{ac}L_{bd}- \delta_{bc}L_{ad}-\delta_{ad}L_{bc}).
\label{q5}
\end{equation}
We note that the action of $L_{ab}$ on any 
N-dimensional vector $s_{a}$ with $a=1,...,N$ is given by
\begin{equation}
\left[L_{ab},s_{c}\right] = i(\delta_{ac}s_{b}
-\delta_{bc}s_{a}).
\label{q5a}
\end{equation}
For example the familiar [rotation group in 3-d space] 
SO(3) has three generators. For SO(5) there are ten
generators. These generators [matrices] will act on
the 5-d vector $s_{a}$ with $a=1,..,5$, where $s_1$
and $s_5$ are the SC part of the 5-d SO(5) vector and
can be readily identified with d-wave SC-order parameters,
viz,
\begin{eqnarray}
s_{1} &=& \Delta+\Delta^{\dagger},\nonumber\\   
s_{5} &=& i(\Delta-\Delta^{\dagger}),\nonumber\\   
\Delta &=& -i\frac{1}{4}
\sum_{k} |g({\bf k})| (g({\bf k})^{2})^{Q}
c_{{\bf k},i}(\sigma^{y})_{ij}c_{-{\bf k},j},\nonumber\\ 
g({\bf k}) &=& \cos(k_x)-\cos(k_y),
\label{q6}
\end{eqnarray}
and $s_2$, $s_3$ and $s_4$ which represent the AF part
are given by 
\begin{eqnarray}
s_2 &=& -\frac{1}{2} (-1)^{Q}
\sum_{k} (g({\bf k})^{2})^{Q-1}c_{{\bf k+A},i}^{\dagger}
(\sigma^{x})_{ij}c_{{\bf k},j},\nonumber\\
s_3 &=& -\frac{1}{2} (-1)^{Q}
\sum_{k} (g({\bf k})^{2})^{Q-1}c_{{\bf k+A},i}^{\dagger}
(\sigma^{y})_{ij}c_{{\bf k},j},\nonumber\\
s_4 &=& -\frac{1}{2} (-1)^{Q}
\sum_{k} (g({\bf k})^{2})^{Q-1}c_{{\bf k+A},i}^{\dagger}
(\sigma^{z})_{ij}c_{{\bf k},j},\nonumber\\
{\bf A} & = & (\pi,\pi,\pi)
\label{q7}
\end{eqnarray}
where {\bf A} is the AF ordering vector.
So far we have identified four of the ten symmetry
generators of SO(5) group, namely $Q, S_1, S_2, {\rm and}
S_3$. The remaining six can be notated by 
$\Pi_{a}$ [$a=1,2,3$] and $\Pi_{a}^{\dagger}$ and which 
for the quantum deformed SO(5) read as
\begin{eqnarray}
\Pi_1 &=& -\frac{1}{2} (-1)^{Q}
\sum_{k} |g({\bf k})|(g({\bf k})^{2})^{Q}c_{{\bf k+A},i}
(\sigma^{y}\sigma^{x})_{ij}c_{-{\bf k},j},\nonumber\\
\Pi_2 &=& -\frac{1}{2} (-1)^{Q}
\sum_{k} |g({\bf k})|(g({\bf k})^{2})^{Q}c_{{\bf k+A},i}
(\sigma^{y}\sigma^{y})_{ij}c_{-{\bf k},j},\nonumber\\
\Pi_3 &=& -\frac{1}{2} (-1)^{Q}
\sum_{k} |g({\bf k})|(g({\bf k})^{2})^{Q}c_{{\bf k+A},i}
(\sigma^{y}\sigma^{z})_{ij}c_{-{\bf k},j}.
\label{q8}
\end{eqnarray}
Now the ten generators $L_{ab}$ can be easily constructed
from $Q,~~S_1,~~S_2,~~S_3,~~\Pi_{1},~~\Pi_{2},~~\Pi_{3},~~\Pi_{1}^{\dagger}, 
~~\Pi_{2}^{\dagger},~~{\rm and}~~\Pi_{3}^{\dagger}$ and satisfy the relation
Eq.~\ref{q4} as they should. 
An independent calculation which constructs the deformed
SO(5) algebra by Duc and Thang\cite{duc99} provides a
check on our results and also confirms our earlier
conjecture \cite{alam98,alam99-1}.

A very important feature of quantum deforemed SO(5) is that 
$g({\bf k})$ not only appears in the SC order parameter
but also in the AF-order parameters, see Eqs.\ref{q6}-\ref{q7},
unlike classical SO(5) where it only appears in SC order
parameters. In classical SO(5) of Zhang although a unification
of AF and SC is posited, it does not imply that the pairing
mechanism is AF fluctuations. Moreover classical SO(5) lacks
a mechanism for psuedogap besides other problem. In contrast
SO(5) quantum deformed algebra relates the AF and SC in 
non-trivial ways, for example, $g({\bf k})$ appears in both
AF and SC parameters. Plus one can go further and boldly
suggest that psuedogap region is a consequence of the
competition between the AF and SC orders. The psuedogap
may be related to short-range AF correlations. Since the deformation
parameter enters into the fundamental anticommutation
relations Eq.~\ref{q1} one can think of quantum symmetry
groups as a classification scheme for the various physical
phenomenona in strongly correlated systems, such as 
fractionalization of electron, Luttinger liquid behaviour,
psuedogap and the deviations [anomalies] of Fermi surface 
away from its normal form due to Luttinger Liquid behaviour. 
In a more specific scenario we can think of the
deformation of anticommutation relations Eq.~\ref{q1} as
representative of the correlation-renormalization dressing
effects of fermionic operators.
Thus we can think in a general sense that the 
quantum symmetry could provide a classification of 
strongly correlated behaviour of Mott Insulators. 
Yet another interesting feature of quantum groups is its
relation to various topological orders, this has not
been exploited.

	We note that in order to 
close the SO(5) algebra i.e. to satisfy Eq.~\ref{q4}, Zhang
\cite{zhang97} imposed the condition $g^{2}(k)=1$.
Henley\cite{hen98} made a very useful observation
that by taking $g({\bf k})={\rm sgn}(\cos(k_x)-cos(k_y))$ one
can close the SO(5) algebra without invoking the
adhoc restriction of Zhang\cite{zhang97}. 
Yet another very significant observation is that
d-wave SC {\em does not} follow from the classical
SO(5) of Zhang\cite{zhang97} since one other
inversion-symmetric choice $g({\bf k})={\rm sgn}(\cos(k_x)+cos(k_y))$
is also enough to obtain the closure of the SO(5) algebra
\footnote{This was also realized by Henley\cite{hen98}, ``Interestingly
,one other simple, inversion-symmetric choice would also
satisfy the condition (4): $\eta_{\bf k}={\rm sgn}(\cos(k_x)+cos(k_y))$. That
variant of SO(5), which entails ``extended s-wave'' pairing,
appears free from internal contradictions (contrary to
a suggestion in Ref.[1]).'' We note that ``condition (4)''
refers to closure condition in Henley\cite{hen98} notation,
and Ref.[1] in Henley\cite{hen98} is reference \cite{zhang97}
here.}.
\section{Microscopic Hamiltonians: Comments}
What about the role of phonons in quantum group scenario
for HTSC and related materials? In any case, whatever point
of view one adopts, one must answer the question:
What is the precise role of phonons in cuprates? i.e.
do they play any role?, if not, how can we prove that
this is so? To do so, one way is that we must take a look 
at the various microscopic Hamiltonians related to the basic Hubbard 
model which share with it the common feature of having quantum
group symmetries but are not restricted to half-filling and
to 1-d\footnote{In another scenario we can stay with 1-d, but
go to the string formulation \cite{alam01-x,alam01-y}. Indeed
this may be a more fruitful approach as we have reasoned
in \cite{alam99-1,alam01-x,alam01-y} since low-dimensions [i.e.1-d,
and 2-d] seem to play an important role in physics of cuprates
in both normal and SC states from both experimental and theoretical
points of views.}. 
An extended Hubbard Hamiltonian [EHH] [i.e. with phonons]
with generalized quantum group symmetries away from 
half-filling was proposed by Montorsi and Rasetti\cite{mon94}.
Using mathematical arguments [i.e. Hopf algebra] Schupp
claimed that the quantum symmetry of EHH in \cite{mon94}
exists only on 1-d lattices and in appropriate approximation
is still restricted to half-filling. To this end Schupp \cite{sch97}
suggested an EHH with full quantum symmetry, having the symmetry group
$SU(2)\times SU(2)/Z_2$, not restricted to half-filling and
1-d lattices. Indeed this Hamiltonian can be regarded as
a realization of our suggestion \cite{alam99-1} to use
the quantum symmetry group SO(4) [since $SU(2)\times SU(2)/Z_2$
is equivalent to SO(4)] for a theory of AF and
SC of cuprates. The EHH in \cite{sch97} [see Eq.~23 in \cite{sch97}]
has six-free parameters, its first three terms make up the
Hubbard Model but not restricted to half-filling. This Hamiltonian
also satisfies one of previous aims, that is to find a non-trivial 
relation [unification] of the t-J and Hubbard models. Indeed
the last term of EHH in \cite{sch97} is like the hopping term
in the t-J model with the hopping strength depending on the
occupation of sites. Moreover and importantly after deformation
this t-J like term is the origin of the non-trivial quantum symmetries
of the EHH in \cite{sch97}. In summary the EHH in \cite{sch97}
is one way of realizing the SO(4) quantum group scenario for
a theory of cuprates suggested in \cite{alam99-1}, plus it also
provides a non-trivial way to relate Hubbard and t-J models.
However much work has to done, namely a definitive answer
on the role of phonons in cuprates, and physical reasoning
behind this.     
\section{Experimental Support for our scenario}
The 1-d behavior is intimately tied to quantum group
symmetry. We thus expect that the physics of 
low-dimensional [1-d and 2-d] systems in particular the 
ones which involve correlated electrons can be understood
in a consistent and elegant manner by using quantum groups. 
It is important to note that the 1-d behavior 
of magnetic fluctuation was predicted by our 
theory \cite{alam99-1} before the experiment.
The fluctuations associated with charge were regarded
as 1-d, whereas magnetic fluctuations were regarded
as 2-d. However it was predicted \cite{alam99-1} and
experimentally shown \cite{moo00} that magnetic
fluctuations are also 1-d. Thus in our scenario
all relevant degrees of freedom are 1-d instead of
quasiparticle in the sense of Landau. In one scenario
of our formulation superconductivity arises as a dressed 
stripe phase. We think it may be possible to write down 
transformations equivalent [such as one writes in 
BCS, i.e. Bogoliubov] which can map the 1-d states onto 
the 2-d and 3-d superconducting phase. Another consequence of
our formulation is the carrier inhomogeneity. We have previously 
emphasized the carrier [electron] inhomogeneity in our modelling of
HTSC materials \cite{alam99-1}. In contrast many
models of HTSC assume that charge carrier introduced by doping
distribute uniformly, leading to an electronically homogeneous
system, as in normal metals. However recent experimental
work \cite{pan01} confirms our intuition, which is encouraging.
This inhomogeneity is expected to be manifested in both
the local density of states spectrum and superconducting gap.
This inhomegeniety we suspect appears in the temperature dependent
XANES spectra of Zn-doped LSCO samples\cite{alam02-2}.
In order to confirm the spin-charge separation and 1-d
behaviour we have also suggested SET based experiments for 
the HTSC and related materials\cite{alam02-1}. 
It is interesting to note that a recent paper on pseudogap and 
quantum-transition phenomenology in HTSC cuprates by Tallon et al.,
\cite{tal02} supports our suggestion that psuedogap may be
related to short-range AF correlations.
\section{Conclusions}	
	In conclusion it has been shown explicitly
that one can indeed construct a closed quantum
group SO(5) algebra which relates elegantly AF
with SC without any adhoc restrictions on the
d-wave factor g$^{2}({\bf k})$, as previously suggested in
\cite{alam98,alam99-1}. 
It is important to note that in terms of physics that 
the arbitariness [freedom] of the d-wave factor g$^{2}$(k) 
is tied to quantum group symmetry whereas in order to recover 
classical SO(5) one must set it to unity in an adhoc manner.
Moreover the d-wave factor appears not only in the
SC order parameter but also in the AF order parameters,
thus relating them in a non-trivial manner.
We have also briefly commented on the realization of our
previous conjecture \cite{alam99-1} in context of SO(4)
quantum group, where phonons can be incorporated at least
for 1-d case \cite{mon94} and beyond 1-d and half-filling 
by using EHH\cite{sch97} in a scenario which relates
t-J, Hubbard and related Hamiltonians via twisting ala
Hopf algebras\cite{sch97}.

\section*{Acknowledgments}
The work of Sher Alam is supported by the Japan Society for
Technology [JST] and partially by System Giken. Sher Alam would 
like to tremendously thank and acknowledge E. Ahmed for 
bringing refs. \cite{sch97,mon94,duc99} to his attention.


\end{document}